\documentclass[journal,twocolumn]{IEEEtran}
\usepackage{epsfig,rotating,setspace,latexsym,amsmath,epsf,amssymb,amsfonts,bm,theorem,subfigure,epstopdf}
\DeclareMathOperator*{\argminA}{arg\,min}
\usepackage{cite}
\usepackage{bbm}
\usepackage{algorithmic,algorithm}
\usepackage{color,booktabs,multirow}

\IEEEoverridecommandlockouts
\allowdisplaybreaks
\bstctlcite{IEEEexample:BSTcontrol}

\begin{document}

	\title{Channel-Aware Adversarial Attacks Against Deep Learning-Based Wireless Signal Classifiers}
	
	\author{\IEEEauthorblockN{Brian Kim, Yalin E. Sagduyu, Kemal Davaslioglu, Tugba Erpek, \\ and Sennur Ulukus}
	\thanks{Brian Kim and Sennur Ulukus are with University of Maryland, College Park, MD, USA; email: \{bkim628, ulukus\}@umd.edu.
	Yalin E. Sagduyu and Kemal Davaslioglu are with Intelligent Automation, a BlueHalo Company, Rockville, MD, USA; email: \{ysagduyu, kdavaslioglu\}@i-a-i.com. 
	Tugba Erpek is with Virginia Tech., Hume Center, Arlington, VA, USA; email: {terpek}@vt.edu.
	This effort is supported by the U. S. Army Research Office under contract W911NF-17-C-0090. The content of the information does not necessarily reflect the position or the policy of the U.S. Government, and no official endorsement should be inferred.
	A preliminary version of the material in this paper has partially appeared in Proceedings of Conference on Information Sciences and Systems (CISS) 2020 \cite{Kim}.}}

\maketitle


\begin{abstract}
This paper presents channel-aware adversarial attacks against deep learning-based wireless signal classifiers. There is a transmitter that transmits signals with different modulation types. A deep neural network is used at each receiver to classify its over-the-air received signals to modulation types. In the meantime, an adversary transmits an adversarial perturbation (subject to a power budget) to fool receivers into making errors in classifying signals that are received as superpositions of transmitted signals and adversarial perturbations. First, these evasion attacks are shown to fail when channels are not considered in designing adversarial perturbations. Then, realistic attacks are presented by considering channel effects from the adversary to each receiver. After showing that a channel-aware attack is selective (i.e., it affects only the receiver whose channel is considered in the perturbation design), a broadcast adversarial attack is presented by crafting a common adversarial perturbation to simultaneously fool classifiers at different receivers. The major vulnerability of modulation classifiers to over-the-air adversarial attacks is shown by accounting for different levels of information available about the channel, the transmitter input, and the classifier model. Finally, a certified defense based on randomized smoothing that augments training data with noise is introduced to make the modulation classifier robust to adversarial perturbations.
\end{abstract}

\begin{IEEEkeywords}
Modulation classification, deep learning, adversarial machine learning, adversarial attack, certified defense.
\end{IEEEkeywords}
 
\section{Introduction}
Built upon recent advances in computational resources and algorithmic designs for deep neural networks, \emph{deep learning} (DL) provides powerful means to learn and adapt to rich data representations  \cite{Goodfellow1}. Spectrum data involved in wireless communications is high dimensional due to the underlying channel, interference and traffic effects, and reflects interactions of network protocols that need to solve complex tasks. Thus, DL has found a broad range of applications in wireless systems, e.g., waveform design \cite{erpek1}, spectrum sensing \cite{Davaslioglu2}, channel estimation \cite{song2021saldr,wang2021compressive,jin2019channel}, and signal classification \cite{Oshea2}.

As adversaries can manipulate the training and testing time inputs of machine learning (ML) algorithms, \emph{adversarial ML} has emerged to study the operation of ML models in the presence of adversaries and support safe adoption of ML to the emerging applications \cite{Vorobeychik1}. In particular, deep neural networks (DNNs) are known to be highly susceptible to even small-scale adversarial attacks, as extensively demonstrated in the computer vision domain \cite{Szegedy1}. Shared and broadcast nature of wireless medium increases the potential for adversaries to tamper with DL-based wireless communication tasks over-the-air. However, it is not readily feasible to translate attack and defense mechanisms from other data domains such as computer vision to wireless communications that feature unique characteristics. First, a wireless adversary cannot directly manipulate input data to a DL algorithm running at a separate target and its manipulation needs to reach the target over-the-air through channel effects. Second, a wireless adversary can attack multiple targets (each with a different channel) simultaneously with a single (omnidirectional) transmission (rather than multiple transmissions that would increase the transmit energy consumed and incur more delay) by exploiting the broadcast nature of wireless communications. 

In this paper, we present realistic wireless attacks built upon adversarial ML and the corresponding defense by accounting for channel and broadcast transmission effects in the algorithm design for adversarial attacks. 
We consider the adversarial attack that corresponds to small modifications of the original input to the DNNs that make the DL algorithms misclassify the input. These small modifications are not just random perturbations added as noise (such as in conventional jamming attacks \cite{xu2005feasibility, sagduyu2010wireless, Sagduyu2008}) but they are carefully designed in a way that changes the decision of the DL algorithm. This attack has been first studied for modulation classifier in \cite{Larsson2} by directly manipulating the receiver's input data at the receiver (corresponding to the additive white Gaussian noise (AWGN) channel environment). In this setting, the perturbation has been selected by the adversary according to a power budget without considering channel effects (beyond AWGN channels) on the received perturbation; therefore, this setting corresponds to the best-case scenario for the adversary but it is not practical for realistic channel environments. However, this approach (also followed in subsequent studies \cite{Flowers1,Bair1,Kokalj1,Kokalj2,Kokalj3, Kokalj4, Ke, Lin,lin2, Gunduz1,Gunduz2}) cannot necessarily ensure that the adversarial perturbation that reaches the receiver has the necessary power and phase characteristics to cause misclassification at the target classifier under realistic wireless channel conditions. In addition, only a single receiver has been considered in these previous studies, and broadcast attacks on multiple receivers and associated channel effects have not been studied yet. In this paper, we show the need for a channel-aware adversarial attack that can simultaneously work against multiple receivers by showing that: i) the design of adversarial perturbations without taking realistic channel effects into account cannot fool a modulation classifier, and ii) an adversarial attack crafted for the channel to a particular receiver is not effective against another receiver with different channel characteristics, i.e., the attack is receiver specific and cannot be used for a broadcast attack that is launched simultaneously against multiple receivers. 

By considering a DNN at each receiver to classify wireless signals to modulation types, we design realistic adversarial attacks in the presence of channel effects from the adversary to each receiver and multiple classifiers at different receivers. For that purpose, we start with a single receiver and determine channel-aware adversarial perturbations (subject to a power constraint) to reduce the accuracy of detecting the modulation type at the receiver. We first propose two white-box attacks, a \emph{targeted attack} with minimum power and a \emph{non-targeted attack}, subject to channel effects known by the adversary. We show that the adversarial attack fails to fool the classifier using limited power if the channel between the adversary and the receiver is not considered when designing the adversarial perturbation. Then, we present algorithms to design adversarial perturbations by accounting for known channel effects and show that the classifier accuracy can be significantly reduced by channel-aware adversarial attacks.

By relaxing the assumption of exact channel information at the adversary, we present a white-box adversarial attack with \emph{limited channel information} available at the adversary. We apply principal component analysis (PCA) to generate adversarial attacks using a lower-dimensional representation of channel distribution. In addition to PCA, we also apply variational autoencoder (VAE) to capture the complexity of underlying data characteristics regarding the transmitter input and the channel. We further relax assumptions regarding the information availability of transmitter input and classifier model, and introduce a \emph{black-box universal adversarial perturbation (UAP) attack} for an adversary with limited information. 

We also introduce a \emph{broadcast} adversarial attack to fool multiple classifiers at different receivers with a single perturbation transmission by leveraging the broadcast nature of wireless communications. A practical scenario for this attack is a user authentication system with multiple signal sensors deployed at different locations. First, we show that the channel-aware adversarial perturbation is inherently selective in the sense that it can fool a target classifier at one receiver (whose channel is used in the attack design) but cannot fool a classifier at another receiver due to different channels experienced at different receivers. By considering channels from the adversary to all receivers jointly, we design the broadcast adversarial perturbation that can simultaneously fool multiple classifiers at different receivers. Using different levels of perturbation budget available relative to noise power, our results illustrate the need to utilize the channel information in designing over-the-air adversarial attacks. 

As a countermeasure, we present a \emph{defense} method to reduce the impact of adversarial perturbations on the classifier performance. Following the \emph{randomized smoothing} approach from \cite{Cohen,Lecuyer19,carlini2019} (previously applied in computer vision), the training data for modulation classifier is augmented with isotropic Gaussian noise to make the trained model robust to adversarial perturbations in test time. We show that this defense is effective in reducing the impact of adversarial attacks on the classifier performance. We further consider \emph{certified defense} to guarantee classifier robustness of the classifier accuracy against adversarial attacks. The classifier is certified by augmenting the received signals with Gaussian noise samples in test time and checking statistical significance of classification results.  

In summary, our contributions are given as follows: 
\begin{itemize}
\item We propose different realistic wireless attacks against a modulation classifier by taking channel effects into account when generating the adversarial perturbations. Furthermore, we provide how to generate adversarial attacks when the adversary has limited information regarding channel, input at the classifier, and architecture of the target classifier.
\item We introduce a wireless broadcast adversarial attack that manipulates multiple classifiers at different receivers with a single perturbation. 
\item We present a defense method to mitigate the effect of adversarial attacks on the classifier.
\end{itemize}

The rest of the paper is organized as follows. Section~\ref{sec:RelatedWork} discusses related work. Section~\ref{sec:SystemModel} describes the system model. Sections~\ref{sec:TargetedAttackWChInfo} and \ref{sec:NontargetedAttackWChInfo} present targeted and non-targeted white-box adversarial attacks, respectively, using channel information. Section~\ref{sec:WhiteBoxWOutChInfo} considers the white-box adversarial attack with limited channel information. Section~\ref{sec:UniversalAttack} presents the UAP attack. Section~\ref{sec:broadcasting} introduces the broadcast adversarial perturbation in the presence of multiple receivers. Section~\ref{sec:defense} describes the defense method to make the classifier robust to adversarial perturbations. Section~\ref{sec:Conclusion} concludes the paper.   

\section{Related Work} \label{sec:RelatedWork}
Adversarial ML has been studied in wireless communications \cite{Sagduyu2020, adesina2020adversarial} with different types of attacks. \emph{Exploratory (inference) attacks} have been considered in \cite{Shi2018, Terpek}, where an adversary builds a DNN to learn the transmission pattern in the channel and jams transmissions that would otherwise be successful. Over-the-air spectrum \emph{poisoning (causative) attacks} have been considered in \cite{YiMilcom2018,Sagduyu1}, where an adversary falsifies a transmitter's spectrum sensing data over the air by transmitting during the spectrum sensing period of the transmitter. Poisoning attacks have been extended to cooperative spectrum sensing \cite{Luo2019, ZluoDefense} and IoT data aggregation \cite{ZluoPartialAttack}. \emph{Trojan attacks} have been studied in \cite{Davaslioglu1} against a modulation classifier, where an adversary slightly manipulates training data by inserting Trojans (in terms of phase shifts) that are later activated in test time. \emph{Membership inference attacks} have been considered in \cite{MIA, MIA2} to learn whether a given data sample has been used during training and thereby gaining information on waveform, channel, and radio hardware characteristics. Adversarial ML has been studied in \cite{Shi2019generative, ShiGANSpoofing} to launch \emph{spoofing attacks} that aim to fool signal authentication systems based on DNNs. As a viable threat for emerging wireless systems, adversarial ML has been also studied to launch attacks against 5G and beyond communications \cite{sagduyubc, shiRL, shiflooding}.

\emph{Adversarial attacks} (a.k.a \emph{evasion attacks}) introduce small modifications to the input data that go into the DNNs such that the DL algorithm cannot reliably classify the input data. \cite{Larsson1} has shown that the end-to-end autoencoder communication systems, proposed in \cite{Oshea1}, are vulnerable to adversarial attacks that increase the block-error-rate at the receiver. Adversarial attacks have been also considered against spectrum sensing \cite{Yalin2019}, power allocation \cite{Larsson3, KimPowerControl}, beam selection \cite{KimSSP}, and MIMO channel state information (CSI) feedback \cite{Liu}. One particular application of DNNs that has gained recent interest in wireless communications is signal classification \cite{Oshea2, Oshea3, Dyspan2019, SoltaniAMC, DeepWiFi,wang2021multi,wang2020automatic,lin2020improved}. Adversarial attacks have been studied against modulation classifiers in \cite{Larsson2} and subsequent studies \cite{Flowers1,Bair1,Kokalj1,Kokalj2,Kokalj3, Kokalj4,  Ke, Lin,lin2}. The targeted fast gradient method (FGM) attack \cite{Kurakin1} has been used in \cite{Larsson2} by enforcing the DNNs to misclassify the input signals to a target label. Here, the target is decided by searching over all possible target labels and selecting the one with the least perturbation required to enforce misclassification. It has been shown that the modulation classifier used in \cite{Oshea2} incurs major errors under adversarial attacks with different power levels used for perturbation signals. The Carlini-Wagner attack \cite{Carlini1} has been considered in \cite{Kokalj2} to perturb RF data points and the corresponding defense mechanism based on pre-trained classifier using an autoencoder is considered in \cite{Kokalj4}. Furthermore, defense mechanisms against adversarial attacks for modulation classifier have been studied in \cite{maroto2021safeamc,zhang2021countermeasures}. The setting has been extended to the use of multiple antennas for transmitting perturbations in \cite{KimMultiple} and the effects of surrogate models trained by the adversary has been studied in \cite{KimICC}. Adversarial attack is also used to defend against an intruder trying to intercept the modulation scheme used at the transmitter in \cite{Gunduz1,Gunduz2}, where the adversarial attack is added to the transmitted signal to fool intruders while maintaining the bit error rate between the transmitter and the receiver. This setting has been used in \cite{Kim5G} to enable covert communications by fooling deep-learning-based signal detectors. 

A typical assumption in previous studies has been that the adversarial perturbations is added to the receiver input without considering channel effects from the adversary to the receiver beyond the AWGN channel environment. However, even a small channel effect (pathloss or fading) would significantly reduce the impact of adversarial perturbation by decreasing the received perturbation power just below the necessary power level or tilting the perturbation to a different direction in vector space such that the adversarial attack fails in changing classification decision over the air. Furthermore, due to the broadcast nature of wireless communications, it is possible for an adversary to attack multiple receivers (each with a different channel from the adversary) simultaneously with a common adversarial perturbation although previous works have only considered attacks on a single receiver. This broadcast perturbation attack consumes less transmit energy and incurs less delay compared to sequentially transmitting multiple perturbations, each targeting a different receiver separately.   

\section{System Model} \label{sec:SystemModel}
\begin{figure}[t]
	\centerline{\includegraphics[width=0.7\linewidth]{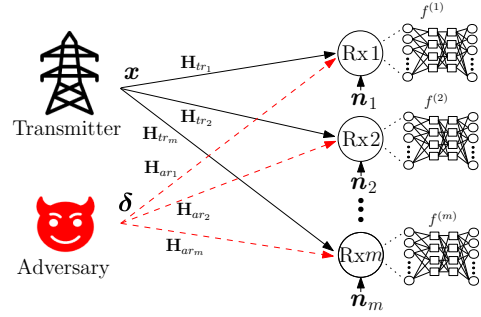}}
	\caption{System model with an adversary transmitting adversarial perturbation to fool multiple receivers.}
	\label{sys}
\end{figure}

We consider a wireless communications system that consists of a transmitter, $m$ receivers, and an adversary as shown in Fig. \ref{sys}. All nodes are equipped with a single antenna and operate on the same channel. Each receiver classifies its received signals with a DNN to the modulation type that is used by the transmitter. In the meantime, the adversary transmits perturbation signals over the air to fool the classifier at the receiver into making errors in modulation classification.  

The DNN classifier at the $i$th receiver is denoted by $f^{(i)}(\cdot;\boldsymbol{\theta}_i): \mathcal{X} \rightarrow \mathbb{R}^{C}$, where $\boldsymbol{\theta}_i$ is the parameters of the DNN at receiver $i$ and $C$ is the number of modulation types. Note $\mathcal{X} \subset \mathbb{C}^{p}$, where $p$ is the dimension of the complex-valued (in-phase/quadrature) inputs that can be also represented by concatenation of two real-valued inputs. The classifier $f^{(i)}$ assigns a modulation type $\hat{l}^{(i)}(\boldsymbol{x},\boldsymbol{\theta}_i) = \arg \max_{k} f^{(i)}_{k}(\boldsymbol{x},\boldsymbol{\theta}_i)$ to every input $\boldsymbol{x}\in \mathcal{X}$ where $f^{(i)}_{k}(\boldsymbol{x},\boldsymbol{\theta}_i)$ is the output of the $i$th classifier corresponding to the $k$th modulation type.

The channel from the transmitter to the $i$th receiver is denoted by $\boldsymbol{h}_{tr_{i}}$ and the channel from the adversary to the $i$th receiver is denoted by $\boldsymbol{h}_{ar_{i}}$, where $\boldsymbol{h}_{tr_{i}} = [h_{tr_{i},1}, h_{tr_{i},2}, \cdots, h_{tr_{i},p}]^{T}\in \mathbb{C}^{p\times 1}$ and $\boldsymbol{h}_{ar_{i}} = [h_{ar_{i},1}, h_{ar_{i},2}, \cdots, h_{ar_{i},p}]^{T}\in \mathbb{C}^{p\times 1}$. Note that $h_{tr_i}$ follows the channel model that is described in \cite{dataset1} and we assume that the channel $h_{ar,j}$ is subject to Rayleigh fading with path-loss and shadowing effects, i.e., ${h}_{ar,j} = K(\frac{d_{0}}{d})^{\gamma}\psi {h}_{ray,j}$ where $K$ is a constant factor, ${h}_{ray,j}$ is Rayleigh fading, $\gamma$ is a path loss exponent, $d_0$ is the reference distance, $d$ is the distance between the adversary and the receiver, and $\psi$ is the shadowing effect to represent the randomness of received signal power in reality. If the transmitter transmits $\boldsymbol{x}$, the $i$th receiver receives $\boldsymbol{r}_{tr_{i}} = \mathbf{H}_{tr_{i}} \boldsymbol{x}+\boldsymbol{n}_{i}$, if there is no adversarial attack, or  receives $\boldsymbol{r}_{ar_{i}} (\boldsymbol{\delta}) = \mathbf{H}_{tr_{i}} \boldsymbol{x}+ \mathbf{H}_{ar_{i}}\boldsymbol{\delta}+\boldsymbol{n}_{i}$, if the adversary launches an adversarial attack by transmitting the perturbation signal $\boldsymbol{\delta}$, where $\mathbf{H}_{tr_{i}} = \mbox{diag} \{h_{tr_{i},1},\cdots, h_{tr_{i},p}\}\in \mathbb{C}^{p\times p}, \mathbf{H}_{ar_{i}} = \mbox{diag}\{h_{ar_{i},1}, \cdots, h_{ar_{i},p}\}\in \mathbb{C}^{p \times p}$, $\boldsymbol{\delta}\in \mathbb{C}^{p\times1}$ and $\boldsymbol{n}_{i}\in \mathbb{C}^{p\times 1}$ is complex Gaussian noise. We assume that the adversarial perturbation $\boldsymbol{\delta}$ is synchronized with the transmitter's signal, thus superimposed with the transmitted signal $\boldsymbol{x}$ at the receiver. To make the attack stealthy  (i.e., hard to detect) and energy efficient, the adversarial perturbation $\boldsymbol{\delta}$ is restricted as $\|\boldsymbol{\delta}\|^{2}_{2}\le P_{\textit{max}}$ for some suitable power budget defined by $P_{\textit{max}}$. 

The adversary determines the (common) adversarial perturbation $\boldsymbol{\delta}$ for the input $\boldsymbol{x}$ and all of the classifiers $f^{(i)}$ for $i = 1,2,\cdots,m$ by solving the  optimization (\ref{broadcasting_eq}) problem:
\begin{align}\label{broadcasting_eq}
\argminA_{\boldsymbol{\delta}} & \quad \; \|\boldsymbol{\delta}\|_{2} \nonumber\\
\mbox{subject to} & \quad \hat{l}^{(i)}(\boldsymbol{r}_{tr_{i}},\boldsymbol{\theta}_i) \ne \hat{l}^{(i)}(\boldsymbol{r}_{ar_{i}}(\boldsymbol{\delta}),\boldsymbol{\theta}_i) \quad i = 1,2, \ldots,m \nonumber \\
& \quad \|\boldsymbol{\delta}\|^{2}_{2} \le P_{\textit{max}}.
\end{align}
In (\ref{broadcasting_eq}), the objective is to minimize the perturbation power subject to two constraints (a) each receiver misclassifies the received signal, and (b) the budget for total perturbation power is not exceeded. Note that the best solution is not necessarily realized at $ \|\boldsymbol{\delta}\|^{2}_{2} = P_{\textit{max}}$ due to the complicated decision boundary of the DNN that depends on the power and phase of perturbation.

In practice, solving (\ref{broadcasting_eq}) is difficult because of the nonlinearity of the DNN. Thus, different methods have been proposed (primarily in the computer vision domain) to approximate the adversarial perturbation. For instance, FGM is a computationally efficient method for crafting adversarial attacks by linearizing the loss function of the DNN classifier. Let $L(\boldsymbol{\theta},\boldsymbol{x},\boldsymbol{y})$  denote the loss function of the model (at any given receiver), where $\boldsymbol{y}\in \{0,1 \}^C$ is the class vector. Then, FGM linearizes the loss function in a neighborhood of $\boldsymbol{x}$ and uses this linearized function for optimization.

We consider two types of attacks called \emph{targeted attacks} and \emph{non-targeted attacks} that involve different objective functions for the adversary to optimize. In a targeted attack, the adversary aims to generate a perturbation that causes the classifier at the receiver to misclassify the input to a target class (label), e.g., a QPSK modulated signal is classified as QAM16. In a non-targeted FGM attack, the adversary searches for an attack that causes any misclassification (independent of target class). More details on these two types of attacks are provided in Section~\ref{sec:TargetedAttackWChInfo}.

Our goal in is paper is to design adversarial perturbation attacks to fool potentially multiple classifiers while considering the channel effects. For the white-box adversarial attacks, we assume that for all $i$ the adversary knows (a) the DNN architecture ($\boldsymbol{\theta}_i$ and $L^{(i)}(\cdot)$) of the classifier at the $i$th receiver, (b) the input at the $i$th receiver, and consequently (c) the channel $\boldsymbol{h}_{ar_{i}}$ between the adversary and the $i$th receiver. We will relax these assumptions in Sections~\ref{sec:WhiteBoxWOutChInfo} and \ref{sec:UniversalAttack}. 

We compare the performances of the attacks proposed in this paper with the benchmark attack from \cite{Larsson2} that generates the adversarial perturbation in the AWGN channel. For performance evaluation, we use the VT-CNN2 classifier from \cite{Oshea1} as the modulation classifier (also used in \cite{Larsson2}), where the classifier consists of two convolution layers and two fully connected layers, and train it with GNU radio ML dataset RML2016.10a \cite{dataset1}. The dataset contains 220,000 samples. Each sample corresponds to a specific modulation scheme at a specific signal-to-noise ratio (SNR). There are eleven modulations in the dataset: BPSK, QPSK, 8PSK, QAM16, QAM64, CPFSK, GFSK, PAM4, WBFM, AM-SSB, and AM-DSB. We follow the same setup of \cite{Oshea1}, using Keras with TensorFlow backend, where the input sample to the modulation classifier is 128 I/Q (in-phase/quadrature) channel symbols. Half of the samples are used for training and the other half are used in test time.

\section{Targeted White-Box Adversarial Attacks using Channel Information} \label{sec:TargetedAttackWChInfo}
We start with a single receiver, i.e., $m=1$, and omit the index $i$ of the $i$th receiver for simplicity in Sections~\ref{sec:TargetedAttackWChInfo}-\ref{sec:UniversalAttack}. We will extend the setup to multiple receivers in Section~\ref{sec:broadcasting}. For the targeted attack, the adversary minimizes $L(\boldsymbol{\theta},\boldsymbol{r}_{ar},\boldsymbol{y}^{\textit{target}})$ with respect to $\boldsymbol{\delta}$, where $\boldsymbol{y}^{\textit{target}}$ is one-hot encoded desired target class (label). FGM is used to linearize the loss function as $L(\boldsymbol{\theta},\boldsymbol{r}_{ar},\boldsymbol{y}^{\textit{target}}) \approx  L(\boldsymbol{\theta},\boldsymbol{r}_{tr},\boldsymbol{y}^{\textit{target}}) + (\mathbf{H}_{ar}\boldsymbol{\delta})^{T} \nabla_{\boldsymbol{x}}L(\boldsymbol{\theta},\boldsymbol{r}_{tr},\boldsymbol{y}^{\textit{target}}) $ that is minimized by setting $\mathbf{H}_{ar}\boldsymbol{\delta} = -\alpha \nabla_{\boldsymbol{x}}L(\boldsymbol{\theta},\boldsymbol{r}_{tr},\boldsymbol{y}^{\textit{target}})$, where $\alpha$ is a scaling factor to constrain the adversarial perturbation power to $P_{\textit{max}}$. 

The adversary can generate different targeted attacks with respect to different $\boldsymbol{y}^{\textit{target}}$ that causes the classifier at the receiver to misclassify the received signals to $C-1$ different modulation types. Thus, as in \cite{Larsson2}, the adversary can craft targeted attacks for all $C-1$ modulation types and choose the target modulation that requires the least power. The case considered in \cite{Larsson2} corresponds to $\mathbf{H}_{ar} = \mathbf{I}$ without realistic channel effects. We call the targeted attack perturbation in \cite{Larsson2} as $\boldsymbol{\delta}^{\textit{NoCh}}$, which is an optimal targeted attack without channel consideration (this corresponds to Algorithm 1 by setting $\mathbf{H}_{ar} = \mathbf{I}$). In the following subsections, we propose three targeted adversarial attacks to account for the effects of the channel. 

\subsection{Channel Inversion Attack}
We first begin with a naive attack, where the adversary designs its attack by inverting the channel in the optimal targeted attack $\boldsymbol{\delta}^{\textit{NoCh}}$, which is obtained using Algorithm 1 with $\mathbf{H}_{ar} = \mathbf{I}$. Since the adversarial attack goes through channel $\boldsymbol{h}_{ar}$,  the $j$th element of the perturbation $\boldsymbol{\delta}$ is simply designed as ${\delta}_{j} = \frac{{\delta}_{j}^{\textit{NoCh}}}{{h}_{ar,j}}$  such that after going through the channel it has the same direction as ${\delta}^{\textit{NoCh}}_{j}$ for $j = 1,\cdots,p$. Furthermore, to satisfy the power constraint $P_{\textit{max}}$ at the adversary, a scaling factor $\alpha$ is introduced such that $\boldsymbol{\delta}^{div} = -\alpha \boldsymbol{\delta}$, where $\alpha = \frac{\sqrt{P_{\textit{max}}}}{\|\boldsymbol{\delta}\|_{2}}$. Thus, the perturbation received at the receiver becomes $\mathbf{H}_{ar}\boldsymbol{\delta}^{div} = -\alpha \boldsymbol{\delta}^{\textit{NoCh}}$.

\subsection{Minimum Mean Squared Error (MMSE) Attack} \label{subsec:TargetedAttackWChInfo_MMSE}
In the MMSE attack, the adversary designs the perturbation $\boldsymbol{\delta}^{\textit{\textit{MMSE}}}$ so that the distance between the perturbation after going through the channel and the optimal targeted perturbation at the receiver (which corresponds to $\mathbf{H}_{ar} = \mathbf{I}$) is minimized. By designing the attack in this way, the received perturbation at the receiver is close to the optimal targeted attack as much as possible while satisfying the power constraint at the adversary. However, since the classifier is sensitive to not only the direction but also to the power of perturbation, the squared error criterion might penalize the candidates of $\boldsymbol{\delta}^{\textit{MMSE}}$ that have more power with the direction of $\boldsymbol{\delta}^{\textit{NoCh}}$, i.e., we set  $\boldsymbol{\delta}^{\textit{MMSE}} = \gamma\boldsymbol{\delta}^{\textit{NoCh}}$ to search for all magnitudes of the $\boldsymbol{\delta}^{\textit{NoCh}}$. Therefore, we formulate the optimization problem to select the perturbation $\boldsymbol{\delta}^{\textit{MMSE}}$ as
\begin{align} \label{prob2}
 \min_{\boldsymbol{\delta}^{\textit{MMSE}}}&  \quad \| \mathbf{H}_{ar}\boldsymbol{\delta}^{\textit{MMSE}} - \gamma\boldsymbol{\delta}^{\textit{NoCh}}\|^{2}_{2}\nonumber\\
 \mbox{subject to}& \quad \|\boldsymbol{\delta}^{\textit{MMSE}} \|^{2}_{2} \le P_{\textit{max}},
\end{align}
where $\gamma$ is optimized by line search. We can write (\ref{prob2}) as
\begin{align} \label{prob3}
	 \min_{{\delta}_{j}^{\textit{MMSE}}} & \quad  \sum^{p}_{j=1} \|h_{ar,j}\delta^{\textit{MMSE}}_{j}-\gamma\delta^{\textit{NoCh}}_{j}\|_2^2             \nonumber\\
	 \mbox{subject to} & \quad   \sum^{p}_{j=1}\|\delta^{\textit{MMSE}}_{j}\|_2^{2}  \le P_{\textit{max}}.
\end{align}
We solve the convex optimization problem (\ref{prob3}) by using the Lagrangian method. The Lagrangian for (\ref{prob3}) is given by
\begin{align}
	\mathcal{L}= & \sum^{p}_{j=1} \|h_{ar,j}\delta^{\textit{MMSE}}_{j}\hspace{-0.5mm}-\hspace{-0.5mm}\gamma\delta^{\textit{NoCh}}_{j}\|_2^2 \hspace{-0.5mm}+\hspace{-0.5mm} \lambda \left( \sum^{p}_{j=1}\|\delta^{\textit{MMSE}}_{j}\|_2^{2}\hspace{-0.8mm} -\hspace{-0.8mm} P_{\textit{max}}\right),
\end{align}
where $\lambda \ge 0$. The Karush–Kuhn–Tucker (KKT) conditions are given by
\begin{align}
	 h^{*}_{ar,j}(h_{ar,j}\delta_{j}^{\textit{MMSE}} - \gamma\delta_{j}^{\textit{NoCh}})+\lambda\delta_{j}^{\textit{MMSE}} = 0 ,
\end{align}  
for $j = 1,\cdots,p$. From the KKT conditions, we obtain the perturbation of the MMSE attack as
\begin{align}
	\delta_{j}^{\textit{MMSE}} = -\frac{\gamma h^{*}_{ar,j}\delta_{j}^{\textit{NoCh}}}{h^{*}_{ar,j}h_{ar,j}+ \lambda},
\end{align} 
for $j = 1,\cdots, p$, where $\lambda$ is determined by the power constraint at the adversary. Note that the received perturbation at the receiver is $\mathbf{H}_{ar}\boldsymbol{\delta}^{\textit{MMSE}} = -\boldsymbol{\alpha}^{T} \boldsymbol{\delta}^{\textit{NoCh}}$, where $\boldsymbol{\alpha} \in \mathbb{R}^{p\times 1}$ and the $j$th element of $\boldsymbol{\alpha}$ is $\alpha_{j} = \frac{\gamma h_{ar,j}h^{*}_{ar,j}}{h^{*}_{ar,j}h_{ar,j}+ \lambda}$. 

\begin{algorithm}[t]
	\setstretch{1.15}
	\caption{MRPP attack}
	\label{target}
	\begin{algorithmic}[1] 
	    \STATE Inputs: input $\boldsymbol{r}_{tr}$, desired accuracy $\varepsilon_{acc}$, power constraint $P_{\textit{max}}$ and model of the classifier
		\STATE Initialize: $ \boldsymbol{\varepsilon}\leftarrow \boldsymbol{0}^{C \times 1}$
		\FOR{class-index $c$ in range($C$)} 
		\STATE $\varepsilon_{\textit{max}} \leftarrow P_{\textit{max}}, \varepsilon_{min} \leftarrow 0$
		\STATE $\boldsymbol{\delta}_{\textit{norm}}^{c} =\frac{\mathbf{H}^{*}_{ar}\nabla_{\boldsymbol{x}}L(\boldsymbol{\theta},\boldsymbol{r}_{tr},\boldsymbol{y}^{c})}{\|\mathbf{H}^{*}_{ar}\nabla_{\boldsymbol{x}}L(\boldsymbol{\theta},\boldsymbol{r}_{tr},\boldsymbol{y}^{c})\|_{2}}$
		\WHILE{$\varepsilon_{\textit{max}}-\varepsilon_{min} > \varepsilon_{acc}$}
		\STATE $\varepsilon_{avg} \leftarrow (\varepsilon_{\textit{max}}+\varepsilon_{min})/2$
		\STATE $\boldsymbol{x}_{adv} \leftarrow \boldsymbol{x} - \varepsilon_{avg}\mathbf{H}_{ar}\boldsymbol{\delta}_{\textit{norm}}^{c}$
		\IF{$\hat{l}(\boldsymbol{x}_{adv})== l_{\textit{true}}$}
		\STATE $\varepsilon_{min}\leftarrow \varepsilon_{avg}$
		\ELSE
		\STATE $\varepsilon_{\textit{max}}\leftarrow \varepsilon_{avg}$
		\ENDIF
		\ENDWHILE
		\STATE $\varepsilon[c] = \varepsilon_{\textit{max}}$
		\ENDFOR 
		\STATE $\textit{target} = \arg\min \boldsymbol{\varepsilon}, \;\boldsymbol{\delta}^{\textit{MRPP}} = -\sqrt{P_{\textit{max}}}\boldsymbol{\delta}_{\textit{norm}}^{\textit{target}}$ 
	\end{algorithmic}
\end{algorithm}

\subsection{Maximum Received Perturbation Power (MRPP) Attack}
In the MRPP attack, the adversary selects the perturbation $\boldsymbol{\delta}^{\textit{MRPP}}$ to maximize the received perturbation power at the receiver and analyzes how the received perturbation power affects the decision process of the classifier. 
To maximize the received perturbation power and effectively fool the classifier into making a specific classification error, the adversary has to fully utilize the channel between the adversary and the receiver. Thus, if the targeted attack ${\delta}^{\textit{target}}_{j}$ is multiplied by the conjugate of the channel, ${h}^{*}_{ar,j}$, to maximize the received perturbation power along the channel ${h}_{ar,j}$, then the received perturbation after going through the channel becomes $\|{h}_{ar,j}\|^{2}_{2}{\delta}^{\textit{target}}_{j}$. In the MRPP attack, not only the direction of the adversarial perturbation is unaffected after going through the channel, but also the power of the adversarial perturbation is maximized by utilizing the channel. Finally, the adversary generates the targeted attack for every possible modulation type to decide the target class and calculate the scaling factor to satisfy the power constraint at the adversary. The details are presented in Algorithm 1.  

\subsection{Attack Performance} \label{subsec:TargetedAttackWChInfo_performance}

We set the parameters of the channel $h_{ar_j}$ as $K = 1, d_{0}=1, d=10, \gamma = 2.7, \psi \sim\ $Lognormal$(0,8)$ and ${h}_{ray,j} \sim \mbox{Rayleigh}(0,1)$, where $\gamma$ is selected to represent urban area cellular radio. We use the perturbation-to-noise ratio (PNR) metric from \cite{Larsson2} that captures the relative perturbation with respect to the noise and measure how the increase in the PNR affects the accuracy of the classifier. Note that as the PNR increases, it is more likely for the attack to be detected by the receiver. In the performance evaluation figures, we denote the targeted attack by `TA', the non-targeted attack by `NTA' and the broadcast attack by `BC'. 

\begin{figure}[t]
	\centerline{\includegraphics[width=0.8\linewidth]{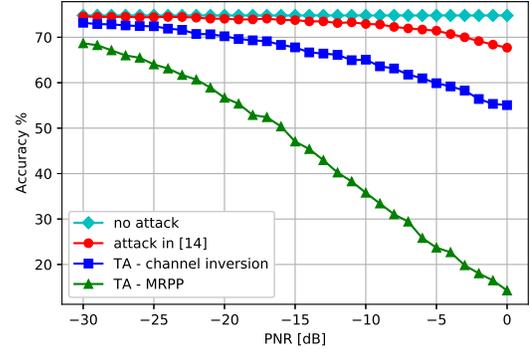}}
	\caption{Accuracy of the receiver's classifier under adversarial attacks with and without considering wireless channel effects when SNR = 10 dB.}
	\label{plot1}
\end{figure}

Fig. \ref{plot1} presents the classifier accuracy versus the PNR under the proposed targeted white-box adversarial attack with exact channel information and compares them with the adversarial attack studied in \cite{Larsson2}. We observe that the white-box attack from \cite{Larsson2} without considering the necessary channel effects from the adversary to the receiver has poor performance that is close to the no attack case in the low PNR region. The reason is that the wireless channel changes the phase and the magnitude of the perturbations perceived at the receiver. Furthermore, the targeted channel inversion attack does not perform well compared to the targeted MRPP attack, indicating the importance of the received power of perturbations for the performance of the classifier at the receiver.

\section{Non-Targeted White-Box Adversarial Attacks using Channel Information} \label{sec:NontargetedAttackWChInfo}
The objective of the non-targeted attack is to maximize the loss function $L(\boldsymbol{\theta},\boldsymbol{r}_{ar},\boldsymbol{y}^{\textit{true}})$, where $\boldsymbol{y}^{\textit{true}}$ is the true class of $\boldsymbol{x}$. FGM is used to linearize the loss function as $L(\boldsymbol{\theta},\boldsymbol{r}_{ar},\boldsymbol{y}^{\textit{true}}) \approx L(\boldsymbol{\theta},\boldsymbol{r}_{tr},\boldsymbol{y}^{\textit{true}}) + (\mathbf{H}_{ar}\boldsymbol{\delta})^{T} \nabla_{\boldsymbol{x}}L(\boldsymbol{\theta},\boldsymbol{r}_{tr},\boldsymbol{y}^{\textit{true}})$ that is maximized by setting $\mathbf{H}_{ar}\boldsymbol{\delta} = \alpha \nabla_{\boldsymbol{x}}L(\boldsymbol{\theta},\boldsymbol{r}_{tr},\boldsymbol{y}^{\textit{true}})$, where $\alpha$ is a scaling factor to constrain the adversarial perturbation power to $P_{\textit{max}}$. Based on FGM, we introduce three non-targeted adversarial attacks described in the following subsections.

\begin{algorithm}[t]
	\setstretch{1.15}
	\caption{Naive non-targeted attack}
	\label{non_target}
	\begin{algorithmic}[1]  
		\STATE Inputs: input $\boldsymbol{r}_{tr}$, number of iterations $E$, power constraint $P_{\textit{max}}$, true class $\boldsymbol{y}^{\textit{true}}$ and model of the classifier
		\STATE Initialize: Sum of gradient $\Delta \leftarrow 0 $ , $\boldsymbol{x}\leftarrow \boldsymbol{r}_{tr}$
		\FOR{$e$ in range($E$)} 
		\STATE $\boldsymbol{\delta}_{\textit{norm}} =\frac{\nabla_{\boldsymbol{x}}L(\boldsymbol{\theta},\boldsymbol{x},\boldsymbol{y}^{\textit{true}})}{\|\nabla_{\boldsymbol{x}}L(\boldsymbol{\theta},\boldsymbol{x},\boldsymbol{y}^{\textit{true}})\|_{2}}$
		\STATE $\boldsymbol{x} \leftarrow \boldsymbol{x} + \sqrt{\frac{{P_{\textit{max}}}}{E}}\mathbf{H}_{ar}\boldsymbol{\delta}_{\textit{norm}}$
		\STATE $\Delta \leftarrow \Delta+ \sqrt{\frac{{P_{\textit{max}}}}{E}}\boldsymbol{\delta}_{\textit{norm}}$
		\ENDFOR 
		\STATE $\boldsymbol{\delta}^{\textit{naive}} = \sqrt{P_{\textit{max}}}\frac{\boldsymbol{\Delta}}{\|\boldsymbol{\Delta}\|_{2}}$		
	\end{algorithmic}
\end{algorithm}

\subsection{Naive Non-Targeted Attack} \label{sec:attack1}
First, the adversary divides its power $P_{\textit{max}}$ into the number of iterations, $E$, in Algorithm 2 and allocates $\frac{{P_{\textit{max}}}}{E}$ amount of power to the gradient of loss function to tilt the transmitted signal from the transmitter. Next, the adversary calculates the gradient again with respect to the transmitted signal from the transmitter and added perturbation. Then, the adversary adds another perturbation with power $\frac{{P_{\textit{max}}}}{E}$ using the new gradient $\boldsymbol{\delta}_{\textit{norm}}$ as
\begin{equation}
    \boldsymbol{x} \leftarrow \boldsymbol{x} + \sqrt{\frac{{P_{\textit{max}}}}{E}}\mathbf{H}_{ar}\boldsymbol{\delta}_{\textit{norm}}.
\end{equation}
This scheme generates the best direction to increase the loss function at that specific instance. Finally, the adversary repeats this procedure $E$ times and sums all the gradients of the loss function that were added to the transmitted signal from the transmitter since the adversary can send only one perturbation at a time over the air. Finally, a scaling factor is introduced to satisfy the power constraint at the adversary. The details are presented in Algorithm \ref{non_target}.

\subsection{Minimum Mean Squared Error (MMSE) Attack} \label{sec:attack2}
The non-targeted MMSE attack is designed similar to the targeted MMSE attack. The adversary first obtains $\boldsymbol{\delta}^{\textit{NoCh}}$ from the naive non-targeted attack with $\mathbf{H}_{ar} = \mathbf{I}$ and uses it to solve problem (\ref{prob2}). Thus, the solution is the same as the solution to (\ref{prob2}) except that it has the opposite direction to maximize the loss function, whereas the targeted attack minimizes the loss function. Therefore, the perturbation selected by the MMSE scheme for the non-targeted attack is $\boldsymbol{\delta}^{\textit{MMSE}} =  \boldsymbol{\alpha}^{T} \boldsymbol{\delta}^{\textit{NoCh}}$, where $\boldsymbol{\alpha} \in \mathbb{R}^{p}$ and the $j$th element of $\boldsymbol{\alpha}$ is $\alpha_{j} = \frac{\gamma h^{*}_{ar,j}}{h^{*}_{ar,i}h_{ar,j}+ \lambda}$.

\subsection{Maximum Received Perturbation Power (MRPP) Attack} \label{sec:attack3}
As we have seen in the targeted MRPP attack, the attack should be in the form of $\boldsymbol{\delta}^{\textit{MRPP}} = \mathbf{H}^{*}_{ar}\boldsymbol{\delta}^{target}$ to maximize the received perturbation power at the receiver. Thus, the naive non-targeted attack is changed to create the MRPP non-targeted attack by replacing $\boldsymbol{\delta}_{\textit{norm}}$ in Line 5 of Algorithm \ref{non_target} with
\begin{align} \label{deltanormwithchannel}
\boldsymbol{\delta}_{\textit{norm}} =  \frac{\mathbf{H}_{ar}^{*}\nabla_{\boldsymbol{x}}L(\boldsymbol{\theta},\boldsymbol{x},\boldsymbol{y}^{\textit{true}})}{\|\mathbf{H}_{ar}^{*}\nabla_{\boldsymbol{x}}L(\boldsymbol{\theta},\boldsymbol{x},\boldsymbol{y}^{\textit{true}})\|_{2}}.  
\end{align}

\subsection{Attack Performance}

\begin{figure}[t]
	\centerline{\includegraphics[width=0.8\linewidth]{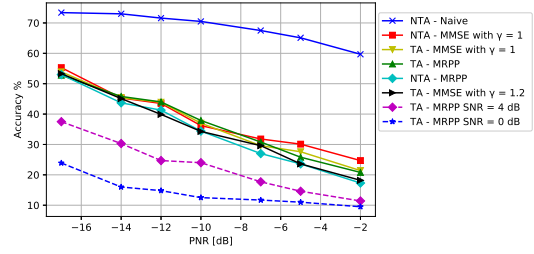}}
	\caption{Accuracy of the receiver's classifier under different white-box adversarial attacks and SNR levels.}
	\label{plot2}
\end{figure}

The performances of the proposed white-box attacks are compared in Fig. \ref{plot2} under the same simulation environment used in Section~\ref{subsec:TargetedAttackWChInfo_performance}. Note that SNR level is set to 10dB if it is not stated in the legend of the figure. As we discussed in Section~\ref{subsec:TargetedAttackWChInfo_MMSE}, $\gamma$ can be optimized with line search for the targeted MMSE attack, e.g., it performs better with $\gamma = 1.2$ compared to $\gamma = 1$. The naive non-targeted attack performs poorly compared to other attacks and the non-targeted MRPP attack outperforms all other attacks {for most of the PNR values}. This can be explained by the freedom of the direction that the non-targeted adversarial attack can take. For targeted attacks, there are only 10 different directions since there are 11 modulation types. However, the non-targeted attack does not have such a restriction. Thus, it is more likely that the non-targeted attack chooses a better direction to enforce a misclassification. Moreover, the computation complexity for the non-targeted attacks is lower compared to the targeted attacks which involve iterations to reach the desired accuracy. Finally, we observe that when the SNR level decreases for the MRPP attack, the accuracy of the classifier also decreases when the same PNR level is considered.

\section{White-box Adversarial Attack with Limited Channel Information} \label{sec:WhiteBoxWOutChInfo}
The adversarial attacks that are designed in the previous sections use the exact information of the channel (from the adversary to the receiver). However, this information may not always be available in practical scenarios. Therefore, in this section, we present Algorithm 3 to generate adversarial attacks with limited channel information, meaning that the adversary only knows the distribution of the channel between the adversary and the receiver and applies dimensionality reduction by accounting for limited channel information. A classical approach for dimensionality reduction is the principal component analysis (PCA) algorithm that was also used in \cite{Larsson2} for the case when the signal is directly manipulated at the receiver. PCA is used to obtain the principal component that has the largest variance, i.e., PCA finds the principal component that provides the most information about the data in a reduced dimension by projecting the data onto it. PCA can be calculated by the eigenvalue decomposition of the data covariance matrix or the singular value decomposition of a data matrix. 

To generate an adversarial attack with limited channel information, the adversary first generates $N$ realizations of the channel from the adversary to the receiver $\{\mathbf{H}_{ar}^{(1)}, \mathbf{H}_{ar}^{(2)}, \cdots, \mathbf{H}_{ar}^{(N)}\}$ which are different from the actual channel $\mathbf{H}_{ar}$. Then, it generates $N$ adversarial attacks using white-box attack algorithms from the previous sections, either targeted or non-targeted, using $N$ realizations of the observed channel and the known input $\boldsymbol{r}_{tr}$ at the classifier. Finally, it stacks $N$ generated adversarial attacks in a matrix and finds the principal component of the matrix to use it as the adversarial attack with limited channel information. The details are presented in Algorithm \ref{limitch}. 
\begin{algorithm}[t]
	\setstretch{1.15}
	\caption{Adversarial attack with limited channel information using PCA}.
	\label{limitch}
	\begin{algorithmic}[1]  
		\STATE Inputs:  $N$ realizations of channels $\{\mathbf{H}_{ar}^{(1)}, \mathbf{H}_{ar}^{(2)}, \cdots, \mathbf{H}_{ar}^{(N)}\}$, input $\boldsymbol{r}_{tr}$ and model of the classifier
		\STATE Initialize: $\Delta \leftarrow 0 $
		\FOR {$n$ in range($N$)} 
		\STATE Find $\boldsymbol{\delta}^{(n)}$ from white-box attack algorithm using $\boldsymbol{r}_{tr}$ and $\mathbf{H}^{(n)}_{ar}$
		\STATE Stack $\boldsymbol{\delta}^{(n)}$ to $\Delta$
		\ENDFOR 
		\STATE Compute the first principle direction $\boldsymbol{v}_{1}$ of $\Delta$ using PCA 
		\STATE $\Delta = \mathbf{U}\mathbf{\Sigma}\mathbf{V}^{T}$ and $\boldsymbol{v}_{1} = \mathbf{V}\boldsymbol{e}_{1}$
		\STATE $\boldsymbol{\delta}^{\textit{limited}} = \sqrt{P_{\textit{max}}}\boldsymbol{v}_{1}$
	\end{algorithmic}
\end{algorithm}

\section{Universal Adversarial Perturbation (UAP) Attack} \label{sec:UniversalAttack}
In the previous sections, the adversary designs a white-box attack with the assumptions that it knows the model of the classifier at the receiver, the channel between the adversary and the receiver, and the exact input at the receiver. However, these assumptions are not always practical in real wireless communications systems. Thus, in this section, we relax these assumptions and present the UAP attacks.

\subsection{UAP Attack with Pre-Collected Input at the Receiver}
We first relax the assumption that the adversary knows the exact input of the classifier. The adversary in the previous attacks generates an input-dependent perturbation, i.e., $\boldsymbol{\delta}$ is designed given the exact input $\boldsymbol{r}_{tr}$. This requires the adversary to always know the input of the classifier, which is not a practical assumption to make and hard to achieve due to synchronization issues. Thus, it is more practical to design an input-independent UAP. We consider two different methods to design an input-independent UAP. We assume that the adversary collects some arbitrary set of inputs $\{\boldsymbol{r}^{(1)}_{tr}, \boldsymbol{r}^{(2)}_{tr},\cdots, \boldsymbol{r}^{(N)}_{tr}\}$ and associated classes.

\subsubsection{PCA-based input-independent UAP attack}
We present an algorithm to design the UAP attack using PCA. First, the adversary generates perturbations $\{\boldsymbol{\delta}^{(1)}, \boldsymbol{\delta}^{(2)},\cdots,  \boldsymbol{\delta}^{(N)}\}$ with respect to the obtained arbitrary set of inputs $\{\boldsymbol{r}^{(1)}_{tr}, \boldsymbol{r}^{(2)}_{tr},\cdots, \boldsymbol{r}^{(N)}_{tr}\}$ and the exact channel information using algorithms from the previous sections such as {the} MRPP attack. To reflect the common characteristics of $\{\boldsymbol{\delta}^{(1)}, \boldsymbol{\delta}^{(2)},\cdots,  \boldsymbol{\delta}^{(N)}\}$ in the UAP, we stack these perturbations into a matrix and perform PCA to find the first component of the matrix with the largest eigenvalue. Hence, we use the direction of the first principal component as the direction of UAP for channel $\mathbf{H}_{ar}$. The algorithm for the UAP with $N$ pre-collected input data is similar to Algorithm \ref{limitch} and the difference is using pre-collected input instead of different realizations of channel.

\subsubsection{VAE-based input-independent UAP attack}
We present an algorithm to design the UAP attack using a VAE that is known to effectively capture complex data characteristics. An autoencoder consists of two DNNs, an encoder to map the input latent space and a decoder to recover the input data from this latent space. VAE extends this setting by describing a probability distribution for each latent attribute instead of providing a single value to describe each latent state attribute. We first collect an arbitrary set of inputs $\{\boldsymbol{r}^{(1)}_{tr}, \boldsymbol{r}^{(2)}_{tr},\cdots, \boldsymbol{r}^{(N)}_{tr}\}$ to create corresponding perturbations $\{\boldsymbol{\delta}^{(1)}, \boldsymbol{\delta}^{(2)},\cdots,  \boldsymbol{\delta}^{(N)}\}$ using Algorithm \ref{target} with $\mathbf{H}_{ar} = \mathbf{I}$. Then, we train the VAE using these perturbations. The encoder structure of the VAE is presented in Table \ref{encoder} and the decoder structure is presented in Table \ref{decoder}. We use 128 number of filters and kernel size of (1,3) for Conv 1 layer and 40 number of filters and kernel size of (2,3) for Conv 2 layer at the encoder. For the deconvolution layers of the decoder, we use 40 number of filters and kernel size of (2,3) for Deconv 1 layer, 128 number of filters and kernel size of (1,3)  for Deconv 2 layer, and one filter and kernel size of (3,3)  for Deconv 3 layer. Since the encoder learns an efficient way to compress the data into the lower-dimensional feature space, we first use the encoder to decrease the dimension of the perturbations. Specifically, we use $k$ number of perturbations $\{\boldsymbol{\delta}^{(1)}, \boldsymbol{\delta}^{(2)},\cdots,  \boldsymbol{\delta}^{(k)}\}$ as inputs to the encoder and get the corresponding outputs $\{\boldsymbol{z}^{(1)}, \boldsymbol{z}^{(2)},\cdots,  \boldsymbol{z}^{(k)}\}$, where $\boldsymbol{z}^{(i)} \in \mathbb{R}^{q\times 1}, i = 1, \cdots, k$. To reflect the common characteristics of $\{\boldsymbol{\delta}^{(1)}, \boldsymbol{\delta}^{(2)},\cdots,  \boldsymbol{\delta}^{(k)}\}$, we average the corresponding outputs $\{\boldsymbol{z}^{(1)}, \boldsymbol{z}^{(2)},\cdots,  \boldsymbol{z}^{(k)}\}$ and use $\boldsymbol{z}^{avg} = \frac{1}{k}\sum_{i=1}^{k}\boldsymbol{z}^{(i)}$ in the decoder to generate $\boldsymbol{\delta}^{avg}$. Finally, we use channel information to design an input-independent UAP as $\boldsymbol{\delta}^{VAE} = \sqrt{P_{\textit{max}}}\frac{\mathbf{H}_{ar}^{*}\boldsymbol{\delta}^{avg}}{\|\mathbf{H}_{ar}^{*}\boldsymbol{\delta}^{avg}\|}$.

\begin{table}[t]
    \centering
	\begin{minipage}{1\columnwidth}
	    \centering
		\caption{Encoder part of the VAE layout }
		\label{encoder}
		\begin{tabular}{l|l|l}
		\toprule
		Layer & Output dimension &  Number of parameters  \\ \cline{1-3}
		Input& 2 $\times$ 128 &   \\
		Conv 1 & 2 $\times$ 128 $\times$128  & 512   \\
		Conv 2& 2$\times$128$\times$40 &  30760 \\
		Flatten &10240& \\
		Dense 1&16 &  163856\\
		Dense 2&4 &  68 
		\\ \bottomrule
	\end{tabular} \\
	\end{minipage}\\
	\begin{minipage}{1\columnwidth}
		\centering
		\caption{Decoder part of the VAE layout}
		\label{decoder}
		\begin{tabular}{l|l|l}
		\toprule
		Layer & Output dimension & Number of parameters  \\ \cline{1-3}
		Input& 4  &    \\
		Dense & 10240 & 51200   \\
		Deconv 1& 2$\times$128$\times$40  & 9640  \\
		Deconv 2&2 $\times$ 128 $\times$128 & 15488\\
		Deconv 3&2$\times$128 & 1153 
		\\ \bottomrule
	\end{tabular} \\
	\end{minipage} 
\end{table}

\subsection{UAP Attack with Limited Channel Information}
Now, we further relax the assumption that the adversary knows the exact channel between the adversary and the receiver and assume that the adversary only knows the distribution of this channel. We consider two different methods to design the UAP attack with limited channel information.

\subsubsection{PCA-based channel-independent UAP attack}
To design the UAP attack when the adversary only knows the distribution of the channel, we first generate random realizations of the channels $\{\mathbf{H}_{ar}^{(1)}, \mathbf{H}_{ar}^{(2)},\cdots, \mathbf{H}_{ar}^{(N)}\}$ from the distribution. Then, we generate $\boldsymbol{\delta}^{(n)}$ using $\boldsymbol{r}_{tr}^{(n)}$ and $\mathbf{H}_{ar}^{(n)}$ for $n = 1,\cdots,N$ instead of using the real channel $\mathbf{H}_{ar}$. We use PCA to find the first component of the matrix and use it as the direction of UAP. The algorithm for PCA-based channel-independent UAP attack is analogous to Algorithm \ref{limitch} except that we have random realizations of channels as opposed to exact channel used in Algorithm \ref{limitch}.

\subsubsection{VAE-based channel-independent UAP attack}
Similar to the PCA-based channel-independent UAP attack, we first generate random realizations of the channel $\{\mathbf{H}_{ar}^{(1)}, \mathbf{H}_{ar}^{(2)},\cdots, \mathbf{H}_{ar}^{(N)}\}$ from the distribution to train the VAE. We use the encoder to capture the common characteristics of the channel by using $k$ number of channels as inputs to the encoder and average the outputs of the encoder to put into the decoder as we have done in the VAE-based input-independent UAP attack to get $\mathbf{H}_{avg}$. Note that we use the same encoder and decoder structure that is used for VAE-based input-independent UAP attack. Furthermore, we use the steps used in the VAE-based input-independent UAP attack to create $\boldsymbol{\delta}^{avg}$ and design $\boldsymbol{\delta}^{VAE} = \sqrt{P_{\textit{max}}}\frac{\mathbf{H}_{avg}^{*}\boldsymbol{\delta}^{avg}}{\|\mathbf{H}_{avg}^{*}\boldsymbol{\delta}^{avg}\|}$.

\subsection{Black-box UAP Attack}
The last assumption that we will relax is the information about the DNN classifier model at the receiver. To relax this assumption, we use the transferability property of adversarial attacks \cite{Transfer1}. This property states that the adversarial attack crafted to fool a specific DNN can also fool other DNNs with different models with high probability. Therefore, the adversary generates UAP attacks using a substitute DNN that is trained using an arbitrary collected dataset and applies them to fool the actual DNN at the receiver. 

\subsection{Attack Performance}
\begin{figure}[t]
	\centerline{\includegraphics[width=1\linewidth]{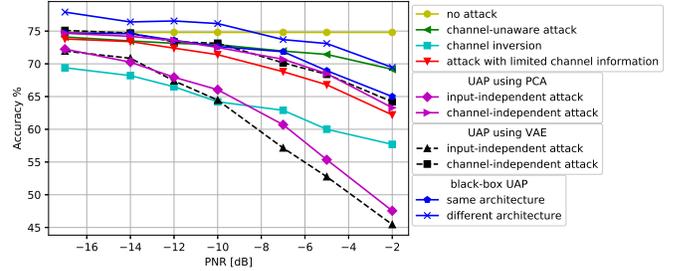}}
	\caption{Accuracy of the receiver's classifier under UAP attacks with different levels of information availability when SNR = 10 dB.}
	\label{plot3}
\end{figure}

Fig.~\ref{plot3} shows the performance of the adversarial attacks with respect to different levels of information availability. First, we observe that the input-independent UAP using VAE with 40 pre-collected data inputs, where the adversary knows the exact channel information, outperforms other attacks with limited information and even the channel inversion attack. Note that VAE is trained with 2000 samples but only 40 samples are used to actually create one UAP. This result shows the importance of the channel state information over the exact input data when crafting an adversarial attack since the input-dependent attack significantly outperforms the attack with limited channel information. Also, the input-independent UAP using VAE outperforms the input-independent UAP using PCA. However, the performance is similar when VAE and PCA are compared for the channel-independent attack suggesting that the UAP using PCA might be enough to generate an effective UAP attack. The input-independent UAP with 40 pre-collected data inputs even outperforms the targeted channel inversion attack in the high PNR region, where the adversary knows not only the exact channel but also the exact input at the receiver. Furthermore, the black-box UAP with the same architecture  performs similar to the attack with limited channel information where we use the substitute DNN that has the same structure of the classifier used in \cite{Oshea1} but is trained with a different training set so that the substitute DNN has different weights. Lastly, when the adversary uses a different DNN architecture that has one more hidden layer with 256 nodes compared to the target DNN model to create an attack, its attack performance is comparable to the channel-unaware attack in the high PNR region.

\section{Broadcast Adversarial Attack}\label{sec:broadcasting}
We extend the adversarial attack to a broadcast (common) adversarial perturbation that is transmitted by the adversary to simultaneously fool multiple classifiers at different receivers. First, we show that the channel-aware adversarial attack built for a particular channel is inherently selective, namely it is only effective against the receiver with that experienced channel and not effective against another receiver with a different experienced channel. Next, we show how to design a common adversarial perturbation by jointly accounting for multiple channels such that the adversary can fool classifiers at multiple receivers simultaneously with a single (omnidirectional) transmission. We present the design of broadcast adversarial attack only for the case of the targeted adversarial attacks, as other types of attacks can be formulated similarly. Note that again we assume the adversary knows the DNN architecture of the receiver, input at the $i$th receiver, and the channel between the adversary and the $i$th receiver in this section.

\subsection{Individual Design of Broadcast Adversarial Attack (IDBA)}
We design IDBA by building upon Algorithm 1 (that was designed against a signal receiver). By applying Algorithm 1 separately against each of $m$ receivers, we obtain $m$ adversarial perturbations, $\boldsymbol{\delta}^{(1)}, \boldsymbol{\delta}^{(2)}, \cdots, \boldsymbol{\delta}^{(m)}$,  where $\boldsymbol{\delta}^{(i)}$ is the adversarial perturbation generated using Algorithm 1 for the $i$th receiver. Since the adversary aims to fool all the receivers by broadcasting a single adversarial perturbation, we combine individual perturbations as a weighted sum, $\sum_{i=1}^m w_i \boldsymbol{\delta}^{(i)}$, where $w_i$ is the weight for adversarial perturbation $\boldsymbol{\delta}^{(i)}$, and then normalize it to satisfy the power constraint at the adversary (higher $w_i$ implies more priority given to $\boldsymbol{\delta}^{(i)}$). Note that the optimal weights can be found by line search subject to $\sum_{i=1}^{m} w_i = 1$ and $w_i  \geq 0$ for each $i$. Numerical results suggest that the search space can be constrained by selecting weights inversely proportional to channel gains of corresponding receivers. 

\subsection{Joint Design of Broadcast Adversarial Attack (JDBA)}
We generate the adversarial perturbation JDBA by modifying Algorithm 1. Specifically, we change the computation of $\boldsymbol{\delta}^{c}_{norm}$ in Line 5 of Algorithm 1 from (\ref{deltanormwithchannel}) to 
\begin{align}
\boldsymbol{\delta}_{\textit{norm}}^{c} =\frac{\sum_{i=1}^{m} w_i\mathbf{H}^{(i)*}_{ar}\nabla_{\boldsymbol{x}}L_{i}(\boldsymbol{\theta},\boldsymbol{r}^{(i)}_{tr},\boldsymbol{y}^{c})}{\|\sum_{i=1}^{m}w_i\mathbf{H}^{(i)*}_{ar}\nabla_{\boldsymbol{x}}L_{i}(\boldsymbol{\theta},\boldsymbol{r}^{(i)}_{tr},\boldsymbol{y}^{c})\|_{2}},
\end{align}
search over all $C-1$ classes, and choose the class that fools most of the receivers. Note that JDBA searches for a common direction of $\boldsymbol{\delta}^{c}_{norm}$ for all receivers. On the other hand, IDBA searches separately for directions of adversarial perturbations for different receivers and then combines them to a common direction.

\subsection{Attack Performance}
For performance evaluation, we assume that there are two receivers, $m=2$, with different classifiers which have the same architecture, but trained with different training sets and the adversary generates a broadcast adversarial perturbation to fool both receivers simultaneously. We assume Rayleigh fading for both channel ${h}_{ar_{1},j} = K(\frac{d_{0}}{d})^{\gamma}\psi  {h}_{ray_{1},j}$ from the adversary to receiver 1, and the channel ${h}_{ar_{2},j} = K(\frac{d_{0}}{d})^{\gamma}\psi {h}_{ray_{2},j}$ from the adversary to receiver 2, where ${h}_{ray_{1},j}$ and ${h}_{ray_{2},j} \sim \mbox{Rayleigh}(0,1)$.

\begin{figure}[t]
	\centerline{\includegraphics[width=0.8\linewidth]{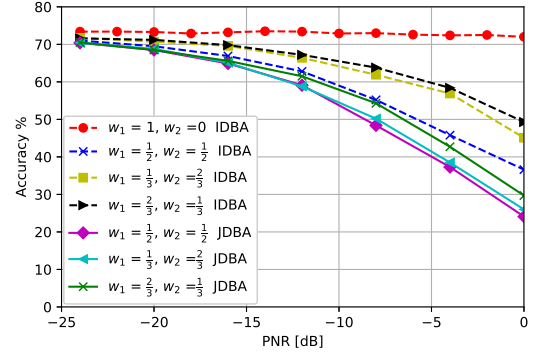}}
	\caption{Accuracy of $m=2$ classifiers under broadcast adversarial attacks when SNR = 10 dB.}
	\label{plot4}
\end{figure}

Fig. \ref{plot4} shows the accuracy of the classifiers with respect to different broadcast adversarial attacks. Note that the accuracy is the same for both classifiers since we assume that the broadcast adversarial attack is successful if it fools both receivers at the same time.  JDBA outperforms IDBA for all PNRs. Furthermore, the weight selection $w_1 = \frac{1}{2}, w_2=\frac{1}{2}$ outperforms other weight selections suggesting that the weight selection is related to the channel distribution. Also, when weights $w_1= 1$ and $w_2= 0$ are selected in IDBA, the attack generated only for receiver 1 has no effect on receiver 2, i.e., the attack selectively fools receiver 1 due to different channel effects. This observation validates the need to craft a broadcast adversarial perturbation.

\begin{figure}[t]
	\centerline{\includegraphics[width=0.8\linewidth]{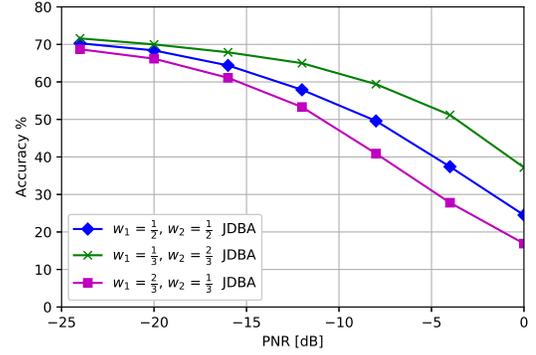}}
	\caption{Accuracy of $m=2$ classifiers under broadcast adversarial attacks at receivers with different channel distributions when SNR = 10 dB.}
	\label{plot5}
\end{figure}

Next, we distinguish channel distributions for the two receivers and evaluate the classifier accuracy under the JDBA attack in Fig. \ref{plot5}. The channel from the adversary to receiver 1, and the channel from the adversary to receiver 2 are modeled with Rayleigh fading of different variances, i.e., ${h}_{ray_1,j}\sim \mbox{Rayleigh}(0,1)$ and ${h}_{ray_2,j} \sim \mbox{Rayleigh}(0,2)$. The weight selection $w_1 = \frac{2}{3}$ and $w_2= \frac{1}{3}$ (where the weights are selected as inversely proportional to the variance of the channel as we have seen in Fig. \ref{plot4}) outperforms other weight selections.

\section{Defense Against Adversarial Attacks}\label{sec:defense}
In this section, we introduce a defense method to increase the robustness of wireless signal classification models against adversarial perturbations. There are several methods developed in the computer vision domain to defend against adversarial attacks. One method is \emph{adversarial training} that calculates an adversarial perturbation and adds the perturbed samples to the training set to increase the model robustness. However, to perform adversarial training, the receiver needs to know what kind of adversarial attack is used at the adversary.
In response, a line of work on \emph{certified defense} has been considered \cite{Cohen,Lecuyer19,carlini2019,Franceschi18}. A classifier is said to be \emph{certifiably robust} if the classifier's prediction of a sample $x$ is constant in a small neighborhood around $x$, often defined by a $\ell_2$ or $\ell_\infty$ ball. \emph{Randomized smoothing} (also referred to as Gaussian smoothing) is a certified defense approach, which augments the training set with Gaussian noise to increase the robustness of the classifier to multiple directions. Recent work has shown that a tight robustness guarantee in $\ell_2$ norm can be achieved with randomized smoothing with Gaussian noise \cite{Cohen}. We assume that the receiver has no information of the adversarial attack that is generated at the adversary. Hence, we apply randomized smoothing as a defense mechanism to make the modulation classifier robust against wireless adversarial perturbations over the air. 

In randomized smoothing, there are two parameters $\sigma$ and $k$ that the defender can control, namely the standard deviation of the Gaussian noise $\sigma$ and the number of noisy samples $k$ added to each training sample $r_i$, i.e., $r_i+n_1, r_i+n_2, \ldots, r_i+n_k$, where $n_j$ is Gaussian noise with variance $\sigma^2$. These two parameters balance the trade-off between the resulting classification accuracy and the robustness against perturbations. Note that unlike the images used as input samples in computer vision, the received signals in wireless communications are already subject to noise, and randomized smoothing slightly increases the noise level. However, as we will see in Section~\ref{sec:defense_evaluation}, as the number of data samples in the training set is less than the number of parameters in the neural network, data augmentation in fact improves the classifier performance and Gaussian smoothing does not cause any degradation. 

\subsection{Randomized Smoothing During Training} \label{sec:defense_evaluation}
\begin{figure}[t]
	\centerline{\includegraphics[width=0.8\linewidth]{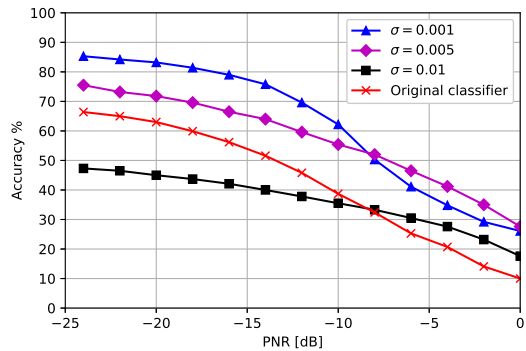}}
	\caption{Accuracy of the receiver's classifier when trained with $(10,\sigma)$ Gaussian noise augmentation with different $\sigma$ and SNR = 10 dB.}
	\label{plot6}
\end{figure}

We evaluate the accuracy of the classifier for different values of $\sigma$ and $k$ selected for Gaussian noise augmentation. First, we fix $k=10$ and change $\sigma$ in Fig. \ref{plot6} to study the impact of the power of the Gaussian noise added to the training set. The accuracy of the classifier trained with small values of $\sigma$ such as $\sigma = 0.001$ and $\sigma = 0.005$ is larger than that of the original classifier across all PNR values. This result shows that randomized smoothing as training data augmentation improves the classifier accuracy for a small $\sigma$ value, while degrading the performance when noise with a large variance is introduced.  

Furthermore, there is an intersection between accuracy versus PNR curves for $\sigma = 0.001$ and $\sigma = 0.005$ due to the density of the norm ball that is created around each training set. These results suggest that when $\sigma$ is small, the classifier becomes more robust to a small perturbation, but more susceptible to a large perturbation. We observe a similar crossover for large $\sigma$ values (e.g., $\sigma=0.01$) where the classifier is more robust to a large perturbation, but susceptible to a small perturbation. 

\begin{figure}[t]
	\centerline{\includegraphics[width=0.8\linewidth]{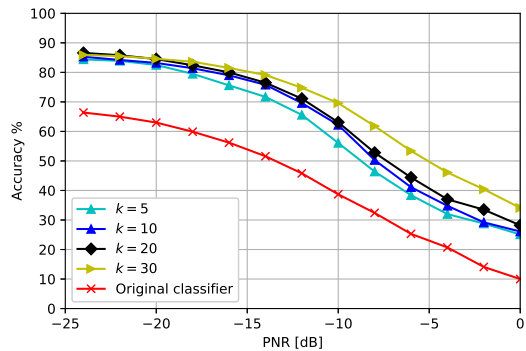}}
	\caption{Accuracy of the receiver's classifier when trained with $(k,0.001)$ Gaussian noise augmentation with different $k$ and SNR = 10 dB.}
	\label{plot7}
\end{figure}

In Fig.~\ref{plot7}, we keep the noise variance $\sigma$ constant and vary the number of samples $k$ added to the training set to investigate its effect on the classifier robustness. Our results demonstrate that by adding more augmentation samples to training set increases the robustness of the classifier against the adversarial perturbation when compared to the original classifier trained without any defense. However, this defense advantage comes at the cost of increased training time.

\subsection{Certified Defense in Test Time}
The defense results can be certified with a desired confidence by using randomized smoothing in test time. Consider the classifier trained with $k=20$ and $\sigma=0.001$. For each perturbed test sample, we draw $k$ Gaussian noise samples with the same variance and label them with the classifier. Let $\hat{c}_A$ and $\hat{c}_B$ denote the classes that occurred most and second most, and $n_A$ and $n_B$ represent the occurrence of these two classes, respectively. We then apply a two-sided hypothesis test and check if $\mathrm{BinomPValue}(n_A,n_A+n_{B},q) \leq \alpha$ condition is satisfied, where $\mathrm{BinomPValue}(\cdot)$ returns $p$-value of the two-sided hypothesis test that $n_A \sim$ Binomial($n_A+n_B,q$) \cite{Cohen}, $1-q$ is the confidence level, and $\alpha$ is the probability of returning an incorrect answer. If the condition is satisfied, the classifier is very confident in its prediction. If not, then the classifier abstains and does not make a prediction. For example, when we consider $95\%$ confidence, we observe that a test sample that belongs to class ``QAM64" is perturbed by the adversary to be classified as ``QAM16" at the receiver. When randomized smoothing is applied in test time, we observe that the classifier infers the samples as class ``QAM16" for 6 times and as class ``QAM64" for 14 times, resulting in an $\mathrm{BinomPValue}(\cdot)=0.115$. Since the condition is not satisfied, the classifier abstains. 

\begin{figure}[t]
\centering`
\subfigure[]{
\includegraphics[width=2.8in,height=2.4in]{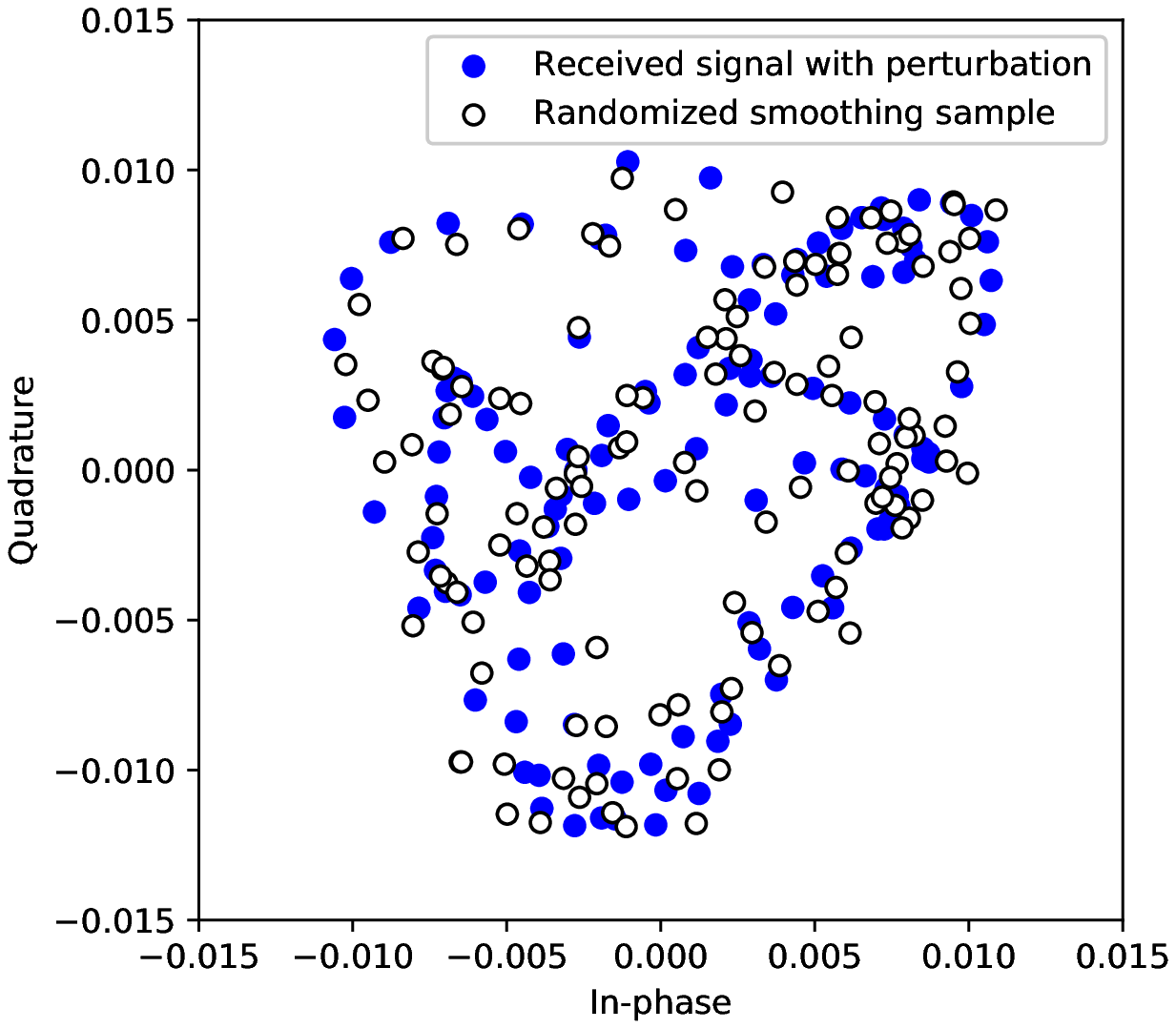}}
\subfigure[]
{\includegraphics[width=2.8in,height=2.4in]{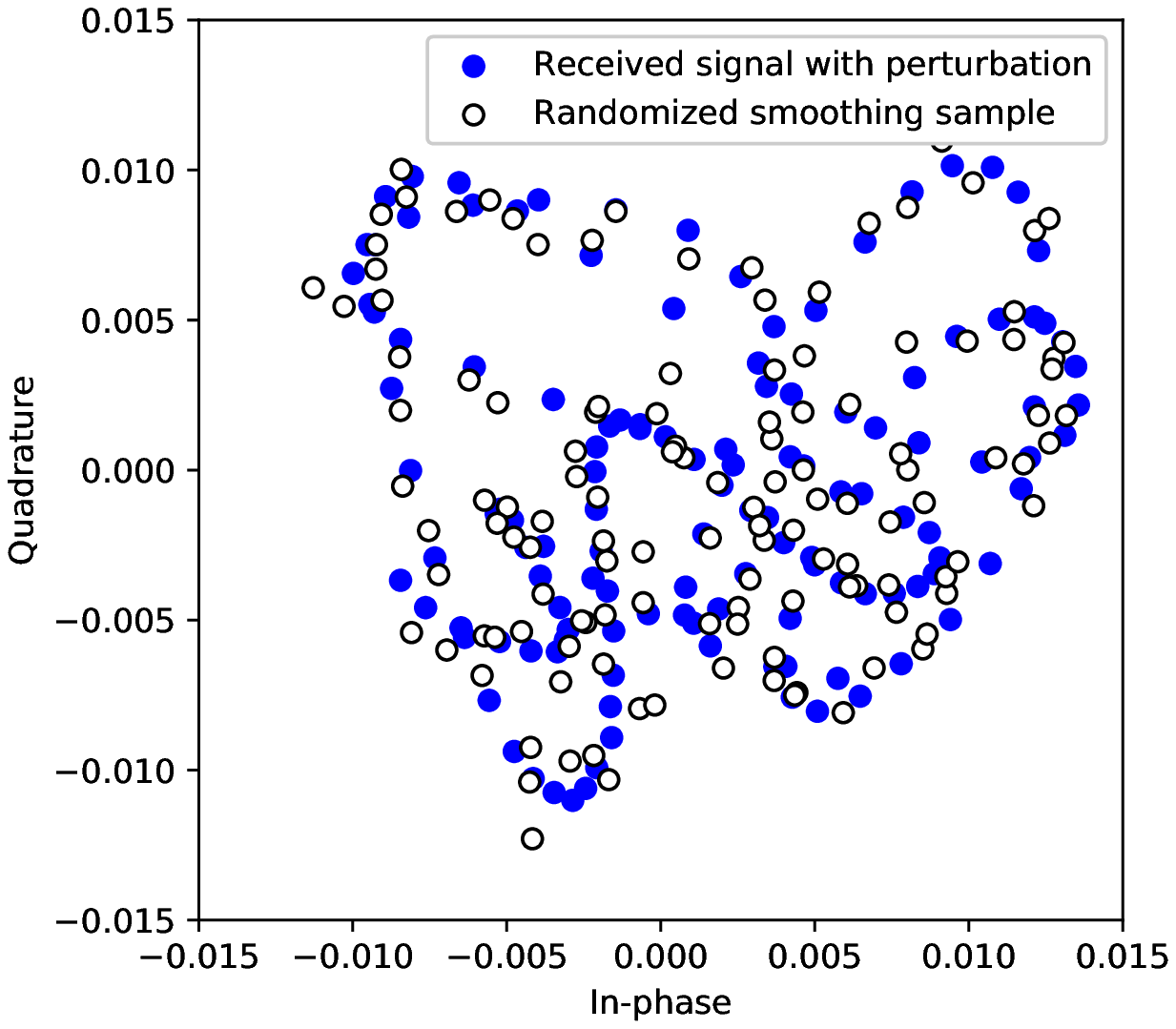}}
\caption{Constellation points of the received signal with adversarial perturbation and randomized smoothing samples for the cases (a) when the classifier abstains and (b) when the classifier recovers the perturbed signal and correctly infers the label.}
\label{fig:Constellation}
\end{figure}

Another example is a test sample of ``QAM64" that is perturbed by the adversary to be classified as ``8PSK". When randomized smoothing is applied in test time, the classifier correctly infers all $k$ samples as ``QAM64" and the defense is certified with $95\%$ confidence. The constellation of the two examples are shown respectively in Fig.~\ref{fig:Constellation}(a)-(b). 

\section{Conclusion} \label{sec:Conclusion}
We presented over-the-air adversarial attacks against deep learning-based modulation classifiers by accounting for realistic channel and broadcast transmission effects. Specifically, we considered targeted, non-targeted and UAP attacks with different levels of uncertainty regarding channels, transmitter inputs and DNN classifier models. We showed that these channel-aware adversarial attacks can successfully fool a modulation classifier over the air. Then, we introduced broadcast adversarial attacks to simultaneously fool multiple classifiers at different receivers with a single perturbation transmission. First, we showed that an adversarial perturbation designed for a particular receiver is not effective against another receiver due to channel differences. Therefore, we designed a common adversarial perturbation by considering all channel effects jointly and showed that this broadcast attack can fool all receivers. Finally, we presented a certified defense method using randomized smoothing, and showed that it is effective in reducing the impact of adversarial attacks on the modulation classifier performance.

\bibliographystyle{ieeetr}
\bibliography{lib}

\begin{IEEEbiography}[{\includegraphics[width=1in,height=1.25in,clip,keepaspectratio]{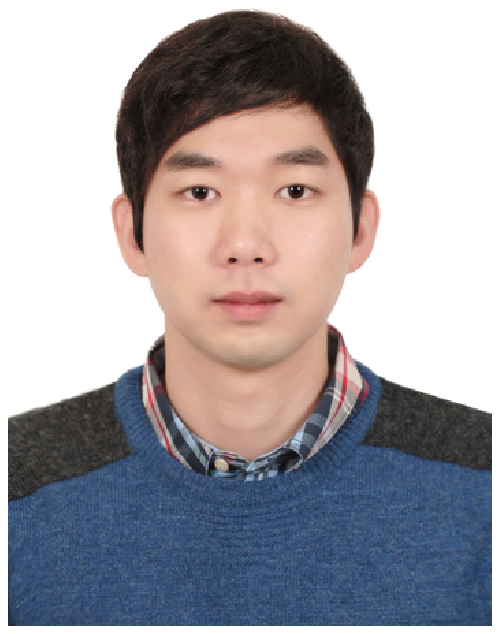}}]{Brian Kim} (Student Member,~IEEE) received the B.S. and M.S. degrees in electrical engineering from Korea Advanced Institute of Science and Technology (KAIST), Daejeon, South Korea, in 2014 and 2016, respectively. At present, he is a Ph.D. candidate in the Department of Electrical and Computer Engineering, University of Maryland at College Park. His research focus is on wireless communications and machine learning.

\end{IEEEbiography}

\begin{IEEEbiography}
[{\includegraphics[width=1in,height=1.25in,clip,keepaspectratio]{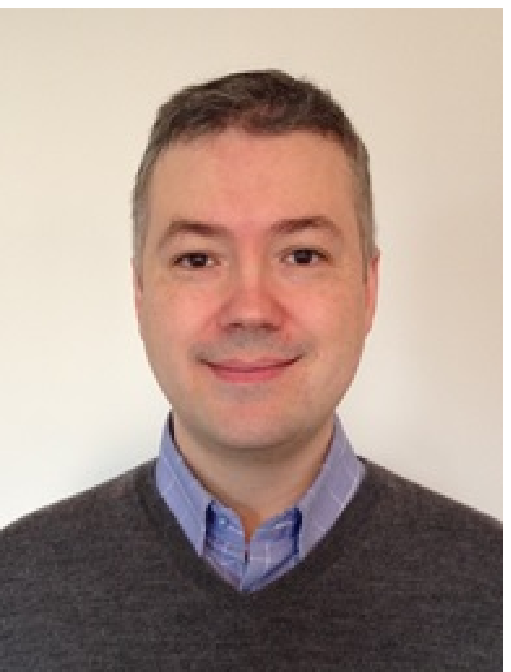}}]
{Yalin E. Sagduyu} (S'02--M'08--SM'15) is the Director of Networks and Security at Intelligent Automation, a BlueHalo Company. He received his Ph.D. degree in Electrical and Computer Engineering from University of Maryland, College Park, in 2007. He has been a Visiting Research Professor in the Department of Electrical and Computer Engineering of University of Maryland, College Park. His research interests are in wireless communications, networks, security, machine learning, 5G and 6G. He chaired workshops at ACM MobiCom, ACM WiSec, IEEE CNS and IEEE ICNP, and served as a Track Chair at IEEE PIMRC, IEEE GlobalSIP and IEEE MILCOM. He received the IEEE HST 2018 Best Paper Award.
\end{IEEEbiography}
\begin{IEEEbiography}[{\includegraphics[height=1.25in,clip,keepaspectratio]{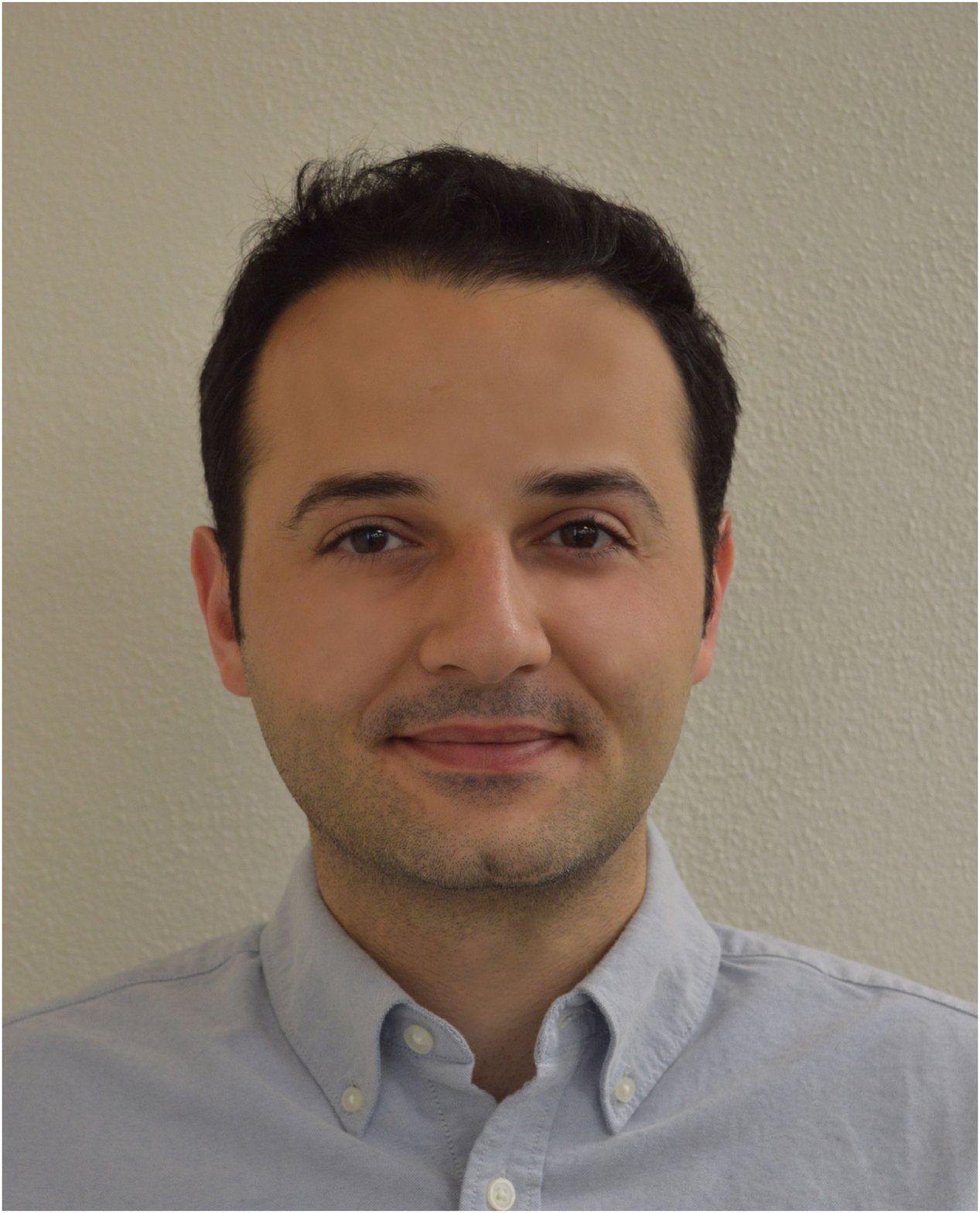}}]{Kemal Davaslioglu} is a Senior Research Scientist at Intelligent Automation, a BlueHalo Company, where he is the Network and Information Systems Technical Area Lead. He received his Ph.D. degree in Electrical and Computer Engineering from University of California, Irvine. Between 2012 and 2015, he held long-term internships at Broadcom Inc., Irvine, CA, where he worked on 10-Gigabit Ethernet systems and later on the beamforming algorithms in the millimeter wave and 60~GHz channel characterization. At Intelligent Automation, a BlueHalo Company, he has developed several deep learning solutions for cognitive radio, adversarial learning, computer vision, synthetic data generation, and cybersecurity. His research interest are on deep learning for wireless networks and computer vision. He has extensive hands-on experience with deep learning software,  its hardware implementation, and software defined radios. He was the recipient of the best paper award at the IEEE Symposium on Technologies for Homeland Security (HST) in 2018. 

\end{IEEEbiography}

\begin{IEEEbiography}[{\includegraphics[width=1.05in,height=1.25in]{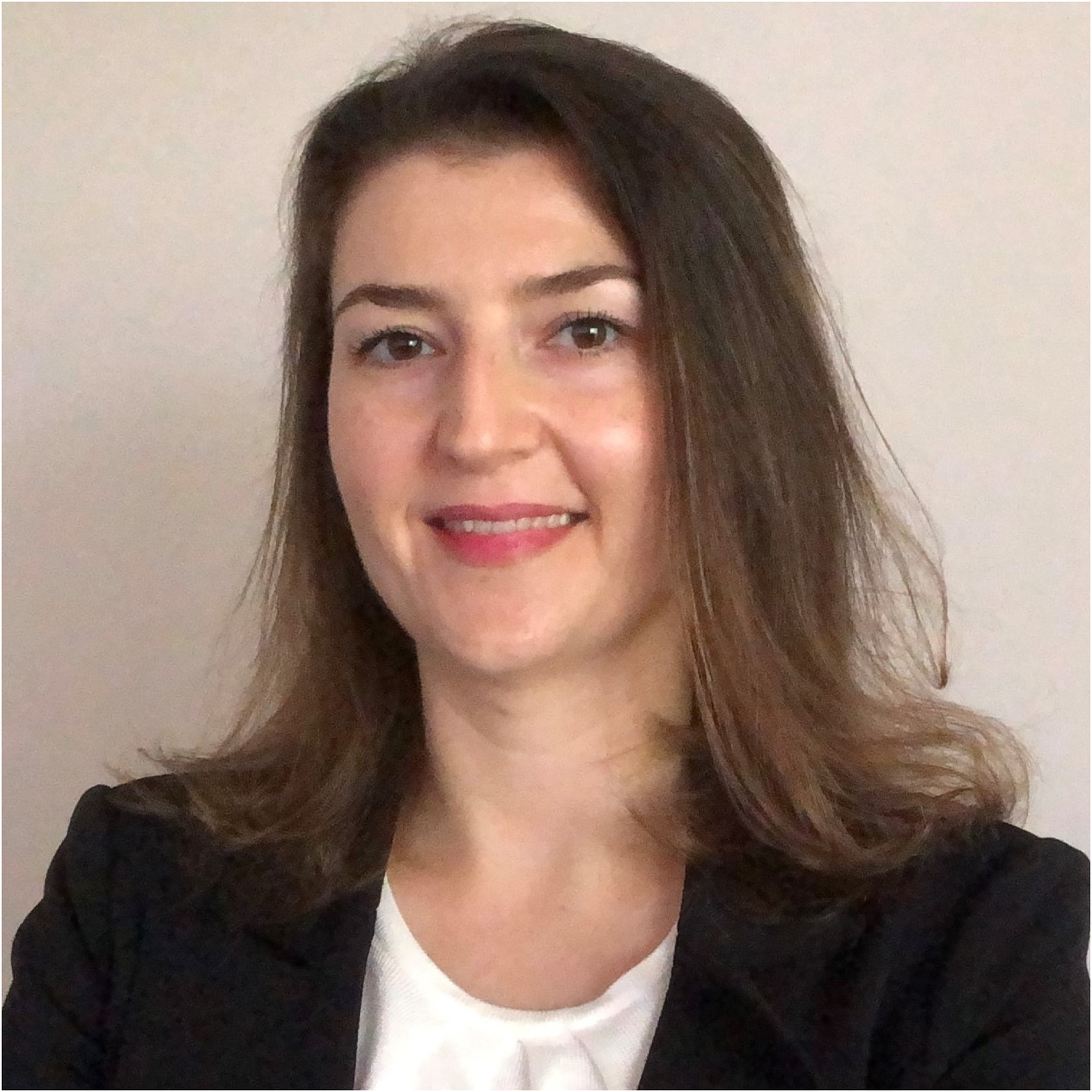}}]{Tugba Erpek} is a Lead Scientist at the Networks and Security Division of Intelligent Automation, a BlueHalo Company, where she is the Network Communications Technical Area Lead. She is also an Adjunct Research Professor at the Hume Center at Virginia Tech. She received her Ph.D. degree in Electrical Engineering from Virginia Tech and has been developing machine/deep learning algorithms to improve the performance, situational awareness, and security of wireless communications systems. Her research interests cover wireless communications and networks, resource allocation, 5G and 6G, physical-layer security, cognitive radio, machine learning, and adversarial machine learning.

\end{IEEEbiography}

\begin{IEEEbiography}[{\includegraphics[width=1in,height=1.25in,clip,keepaspectratio]{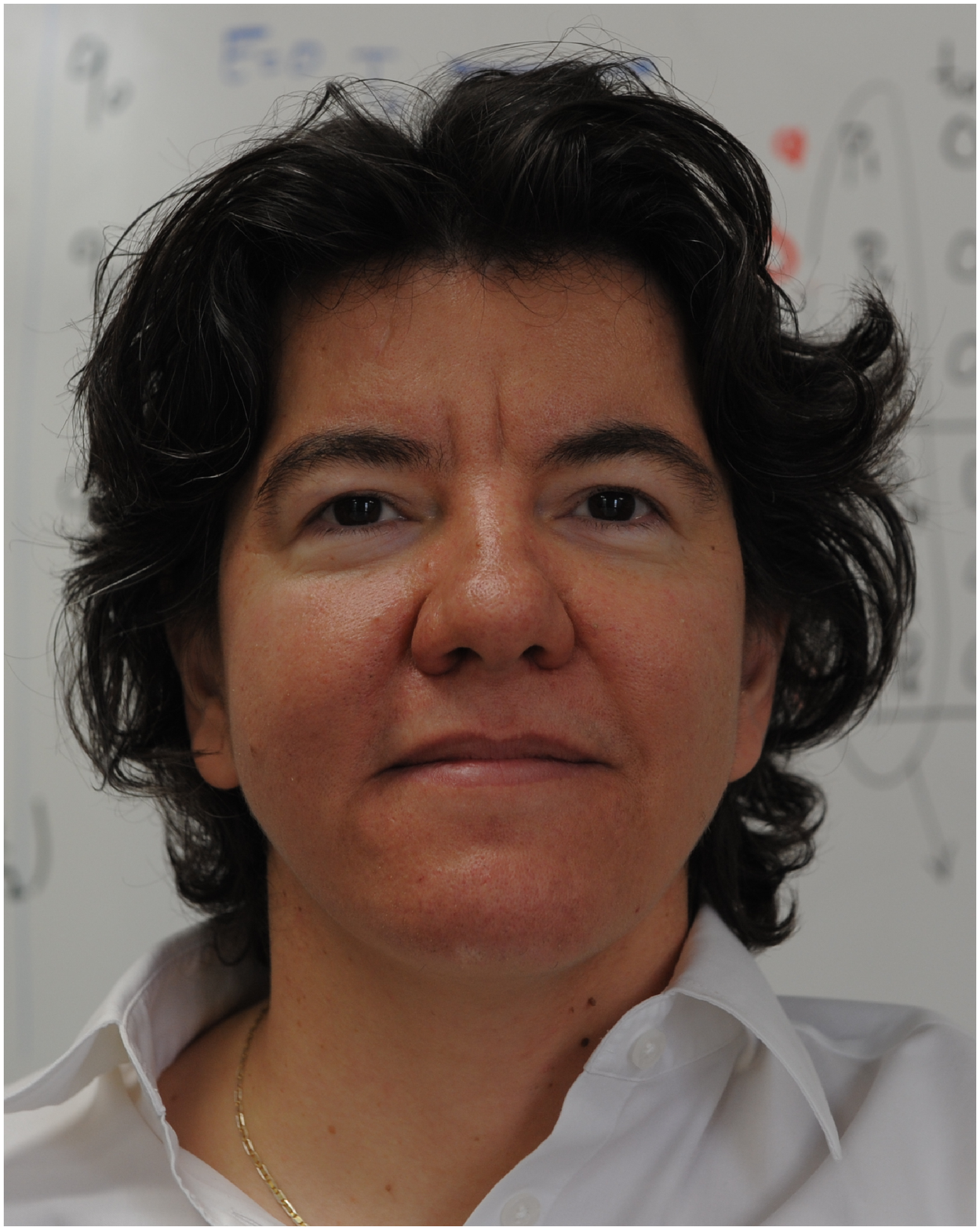}}]{Sennur Ulukus} (Fellow,~IEEE) is the Anthony Ephremides Professor in Information Sciences and Systems in the Department of Electrical and Computer Engineering at the University of Maryland at College Park, where she also holds a joint appointment with the Institute for Systems Research (ISR). Prior to joining UMD, she was a Senior Technical Staff Member at AT\&T Labs-Research. She received her Ph.D. degree in Electrical and Computer Engineering from Wireless Information Network Laboratory (WINLAB), Rutgers University, and B.S. and M.S. degrees in Electrical and Electronics Engineering from Bilkent University. Her research interests are in information theory, wireless communications, machine learning, signal processing and networks, with recent focus on private information retrieval, age of information, group testing, distributed coded computation, energy harvesting communications, physical layer security, and wireless energy and information transfer.

Dr. Ulukus is a fellow of the IEEE, and a Distinguished Scholar-Teacher of the University of Maryland. She received the 2003 IEEE Marconi Prize Paper Award in Wireless Communications, the 2019 IEEE Communications Society Best Tutorial Paper Award, the 2020 IEEE Communications Society Women in Communications Engineering (WICE) Outstanding Achievement Award, the 2020 IEEE Communications Society Technical Committee on Green Communications and Computing (TCGCC) Distinguished Technical Achievement Recognition Award, a 2005 NSF CAREER Award, the 2011 ISR Outstanding Systems Engineering Faculty Award, and the 2012 ECE George Corcoran Outstanding Teaching Award. She was a Distinguished Lecturer of the IEEE Information Theory Society for 2018-2019.

She is an Area Editor for the IEEE Transactions on Wireless Communications (2019-present) and a Senior Editor for the IEEE Transactions on Green Communications and Networking (2020-present). She was an Area Editor for the IEEE Transactions on Green Communications and Networking (2016-2020), an Editor for the IEEE Journal on Selected Areas in Communications-Series on Green Communications and Networking (2015-2016), an Associate Editor for the IEEE Transactions on Information Theory (2007-2010), and an Editor for the IEEE Transactions on Communications (2003-2007). She was a Guest Editor for the IEEE Journal on Selected Areas in Communications (2008, 2015 and 2021), Journal of Communications and Networks (2012), and the IEEE Transactions on Information Theory (2011). She is the TPC chair of 2021 IEEE Globecom, and was a TPC co-chair of 2019 IEEE ITW, 2017 IEEE ISIT, 2016 IEEE Globecom, 2014 IEEE PIMRC, and 2011 IEEE CTW.

\end{IEEEbiography}

\end{document}